\input harvmac

\overfullrule=0pt
\parindent=0pt


\def\NSNS{{$NS\otimes NS$}}
\def\RR{{$R\otimes R$}}
\def\ZZ {{\bf Z}}

\def\det{\hbox{\rm det}}
\def\Sl#1{Sl(#1,\ZZ)}
\def\G(#1){\Gamma(#1)}

\def\C|#1{{\cal #1}}
\def\(#1#2){(\zeta_#1\cdot\zeta_#2)}


\def\xxx#1 {{hep-th/#1}}
\def\lr { \lref}
\def\npb#1(#2)#3 { Nucl. Phys. {\bf B#1} (#2) #3 } \def\rep#1(#2)#3
{ Phys.
Rept.{\bf #1} (#2) #3 } \def\plb#1(#2)#3{Phys. Lett. {\bf #1B} (#2) #3}
\def\prl#1(#2)#3{Phys. Rev. Lett.{\bf #1} (#2) #3}
\def\physrev#1(#2)#3{Phys. Rev. {\bf D#1} (#2) #3} \def\ap#1(#2)#3{Ann.
Phys. {\bf #1} (#2) #3} \def\rmp#1(#2)#3{Rev. Mod. Phys. {\bf #1}
(#2) #3}
\def\cmp#1(#2)#3{Comm. Math. Phys. {\bf #1} (#2) #3}
\def\mpl#1(#2)#3{Mod. Phys. Lett. {\bf #1} (#2) #3}
\def\ijmp#1(#2)#3{Int.
J. Mod. Phys. {\bf A#1} (#2) #3}

\def\calV{{\cal V}}
\def\Rfour{t_8t_8R^4}
\def\hl{\hat l}

\parindent 25pt
\overfullrule=0pt
\tolerance=10000

\sequentialequations



 \lr\schwarz{J.H. Schwarz, {\it An $\Sl2$ multiplet of type IIb
superstrings}, \xxx9508143
\plb360(1995)13.}
\lr\greenschwarzclosed{M.B. Green and J.H. Schwarz, {\it   
Supersymmetrical
string theories}, Phys.  Lett. {\bf 109B} (1982) 444.}
\lr\greenschwarz{M.B. Green and J.H. Schwarz, {\it Supersymmetric  
dual string
theory
(III). Loops and renormalization}, \npb198(1982)441.}
\lr\dkl{L.~Dixon, V.~Kaplunovsky and J.~Louis, {\it Moduli
dependence of String
loop corrections to gauge coupling constants }, \npb355(1991)649}
\lr\greengut{M.B.~Green and M.~Gutperle, {\it Effects of D-instantons},
\xxx9701093.}
\lr\banksetal{   T. Banks, W. Fischler,  S.H. Shenker and L.  
Susskind, {\it M-Theory as a matrix model: a conjecture},
hep-th/9610043,  Phys. Rev. {\bf D55}
(1997) 5112.}
\lr\hulltownsend{C.M.~Hull and P.K.~Townsend, {\it Unity of Superstring
Dualities }, \xxx9410167, \npb438(1995)109.}
\lr\aspinwall{P. Aspinwall, {\it Some Relationships Between
Dualities in String
Theory}, in Proceedings of  \lq S-duality and mirror symmetry', Trieste 1995,
 hep-th/9508154, Nucl. Phys. Proc. {\bf 46} (1996) 30.}
\lr\schwarza {J.H.~Schwarz, {\it Lectures on Superstring and M-theory
dualities},
  \xxx9607201.}
   \lr\luc{M.~L{\"u}scher, K.~Symanzik and P. Weisz, {\it Anomalies  
of the
free loop wave equation in the WKB approximation}, \npb173(1980)365.}
  \lr\julia{ E. Cremmer, B. Julia and J. Scherk, {\it Supergravity
theory in
eleven dimensions}, Phys. Lett. {\bf 76B} (1978) 409.}
\lr\grosswitt{D.J.~Gross and E.~Witten, {\it Superstring modifications of
Einstein's
    equations}, \npb277(1986)1.}
\lr\grisaru{M.T.~Grisaru , A.E.M~Van de Ven and D.~Zanon, {\it
Two-dimensional
supersymmetric sigma models on Ricci flat K{\"a}hler manifolds are not
finite},
\npb277(1986)388 ; {\it Four loop divergences for the N=1 supersymmetric
nonlinear sigma model in two-dimensions}, \npb277(1986)409.}
\lr\polchtasi {J.~Polchinski, {\it TASI Lectures on D-branes},
\xxx9611050.}
\lr\tseytlin{E.S. Fradkin and A.A. Tseytlin, {\it Quantum properties
of higher dimensional and dimensionally reduced supersymmetric theories},
    Nucl. Phys. {\bf B227} (1983) 252.}
\lr\greenhove{M.B. Green and P. Vanhove,  {\it D-instantons, strings and
M-theory},
  hep-th/9704145.}
\lr\terras{A.~Terras, {\it Harmonic Analysis on Symmetric Spaces and
    Applications}, vol.~I, Springer--Verlag (1985).}
\lr\vafwit{C.~Vafa and E.~Witten, {\it A one loop test of
string duality},
  \xxx9505053, \npb447(1995)261.}
\lr\duffliu{M.J.~Duff, J.T.~Liu and R.~Minasian, {\it
Eleven-dimensional
    origin of string-string duality: a one loop test}, \xxx9506126,
  \npb452(1995)261.}
\lr\douglasetal{M.R.~Douglas, D.~Kabat, P.~Pouliot and S.~Shenker, {\it
D-branes and short distances in string theory}, \xxx9608024,
\npb485(1997)85.}
  \lr\lerche{W.  Lerche, {\it Elliptic index and superstring
effective actions},
Nucl. Phys. {\bf B308} (1988) 102.}
\lr\distler{D.~Berenstein, R.~Corrado and J.~Distler, {\it On the moduli space
    of M(atrix)-theory compactification}, hep-th/9704087.}
\lr\ganor{O.J.~Ganor, S.~Ramgoolam and W.~Taylor~IV {\it Branes, fluxes and
    duality in M(atrix)-theory } \xxx9611202}
\lr\greeks{C.~Bachas, C.~Fabre, E.~Kiritsis, N.A.~Obers and P.~Vanhove
  {\it Heterotic/ type-I duality, D-brane instantons and generalized
    prepotentials} (to appear). }
\lr\backir{C.~Bachas and E.~Kiritsis, {\it $F^4$ Terms in N=4 String Vacua},
 Proceedings of Trieste Spring School and Workshop, April 1996
    \xxx9611205.} 
\lr\kallosh{R.  Kallosh, {\it Covariant quantization of D-branes}, hep-th/9705056.}

\noblackbox
\baselineskip 14pt plus 2pt minus 2pt
\Title{\vbox{\baselineskip12pt
\hbox{hep-th/9706175}
\hbox{DAMTP/97-58 }
\hbox{CPTH-S-528-0697 }
}}
{\vbox{
\centerline{ONE LOOP IN ELEVEN DIMENSIONS} }}

\centerline{ Michael B. Green\foot{M.B.Green@damtp.cam.ac.uk}, Michael
Gutperle\foot{ M.Gutperle@damtp.cam.ac.uk},}
\medskip
\centerline{DAMTP, Silver Street, Cambridge CB3 9EW, UK}
 \medskip
\centerline{and}
\medskip
\centerline{Pierre Vanhove\foot{ vanhove@cpth.polytechnique.fr},}
\medskip
\centerline{Centre de Physique Th{\'e}orique, } \centerline{Ecole
Polytechnique, 91128 Palaiseau, FRANCE}

\bigskip\bigskip

\medskip
\centerline{{\bf Abstract}}

Four-graviton scattering in eleven-dimensional supergravity is  
considered at
one loop compactified on one,  two and three-dimensional  tori.   
The dependence
on the toroidal geometry determines  the known perturbative and
non-perturbative terms in the corresponding processes in type II  
superstring
theories  in  ten,  nine and eight dimensions.   The
ultra-violet divergence must be regularized  so that it has  a precisely
determined finite value that is consistent both  with   T-duality   
in nine dimensions and with eleven-dimensional supersymmetry.

\noblackbox
\baselineskip 14pt plus 2pt minus 2pt

\Date{1997}

\newsec{Introduction}

The leading term in the M-theory effective action is the  classical
eleven-dimensional supergravity   of   \refs{\julia}.   Although terms of
higher dimension   must be strongly constrained by the large amount of
supersymmetry  they  have not been systematically investigated.    
There is
known to be an eleven-form, $\int C^{(3)} \wedge X_8$ (where $X_8$ is an
eight-form made out of the curvature $R$ and $C^{(3)}$ is the three-form
potential),  which is necessary for consistency with anomaly cancellation
\refs{\vafwit,\duffliu}. Eleven-dimensional supersymmetry relates this to a 
particular $R^4$ term \refs{\greenhove} as well as a host of other terms and
might  well determine the complete effective action.   Furthermore, 
the effective 
action of the compactified theory  depends nontrivially on moduli fields 
associated with the geometry of
the compact manifold.  This dependence is very strongly restricted by 
consistency with the duality symmetries of string theory in ten and lower 
dimensions.   For example,  the $R^4$ 
term in  M-theory compactified on  a two-torus must be
consistent with the structure of  perturbative and non-perturbative terms in  
nine-dimensional IIA
and IIB superstrings \refs{\greengut}.
This provides  strong evidence that it has  the form \refs{\greenhove}
\eqn\actrfour{S_{R^4} = {1\over  \kappa_{11}^{2/9} }\int  d^9x
\sqrt{ G^{(9)}}h(\Omega, \bar \Omega;{\cal V}_2)  t_8t_8 R^4,}
where $G_{\mu\nu}^{(9)}$ is the   M-theory metric in the space  
transverse to
the torus, $\Omega= \Omega_1 + i \Omega_2$ is the complex structure  
of the
torus and $\calV_2$ is its volume in eleven-dimensional Planck units.  The
 parameter $\kappa_{11}$ has dimensions $[length]^{9/2}$ in arbitrary 
units (and
the volume of the torus is given by $(\kappa_{11})^{4/9} \calV_2$ in these
units).
 The notation $\Rfour$  (to be  
defined below)
indicates the particular contraction  of four Riemann tensors that  
arises from
integration over fermionic zero modes at one loop in superstring theory
\refs{\greenschwarzclosed} as well as from integration over  
fermionic modes on
a D-instanton \refs{\greengut}.  The function $h$ has to be  
invariant under the
action of the modular group, $\Sl2$,  acting on $\Omega$ and in
\refs{\greengut,\greenhove} various arguments were given for why it  
should have
the form
\eqn\hdef{ h(\Omega, \bar \Omega;{\cal V}_2 ) = \calV_2^{-1/2}
f(\Omega,\bar\Omega) + {2\pi^2 \over 3} \calV_2.}
The function $f$  is a modular-invariant non-holomorphic Eisenstein  
series
which is uniquely specified by the fact that it is an eigenfunction  
of the
Laplace operator on the fundamental domain of $\Sl2$ with  
eigenvalue $3/4$,
\eqn\fdef{\Omega_2^2 \ \partial_\Omega \partial_{\bar \Omega} f =  
{3\over 4}\
f}
(i.e.  $f=\zeta(3) E_{3\over 2}$, where $E_s$ is a Maass waveform  
of eigenvalue
$s(s-1)$ \refs{\terras}).   Significantly, this is the kind of Ward  
identity that
the threshold corrections in lower-dimensional $N=2$ theories satisfy
\refs{\dkl} and it suggests a very stringent set of geometrical  
constraints.
The expansion of $f$  for large $\Omega_2$  has two
power-behaved terms plus an infinite series of exponentially  
decreasing terms.
These have exactyl the correct coefficients to be identified 
with the tree-level and one-loop terms,   
together with an
infinite series of D-instanton contributions  in type IIB  
superstring theory.
 This identification makes use of T-duality to relate a multiply-charged  
D-instanton of type
IIB with a multiply-wound world-line of a multiply-charged  
D-particle of type
IIA.  Indeed, semiclassical quantization around these D-instanton
configurations may be carried out by  functional integration around the
D-particle background, as  outlined in \refs{\greenhove}.

In this paper we will show how the sum of the perturbative and   
D-instanton
contributions to the $\Rfour$ term are efficiently encoded in M-theory  
in the
expression for the scattering of four gravitons at one loop in eleven-dimensional supergravity perturbation theory.    The particles circulating around the loop  
are the 256
physical states that comprise the massless  eleven-dimensional  
supergraviton.
It may seem  surprising that perturbation theory  
is of any
significance since supergravity has terrible ultra-violet  
divergences in eleven dimensions.   Furthermore, the absence of any scalar fields  means that there is
no small dimensionless coupling constant.  However, there is strong  
reason to
believe that the one-loop $\Rfour$ terms are protected  from receiving
higher-loop contributions by an eleven-dimensional  
nonrenormalization theorem
since they are related  by supersymmetry to the $C^{(3)} \wedge  
X_8$ term.
Upon compactification to ten or fewer dimensions the Kaluza--Klein  
modes of the
circulating  fields are  reinterpreted in terms of the windings of   
euclidean
D-particle world-lines.   The massive D-particles reproduce the  
D-instanton
effects while the massless one  (the massless ten-dimensional  
supergraviton)
is   equivalent to the perturbative  one-loop string effects.  The  
$\Rfour$
term obtained at tree level in string theory arises, somewhat  
miraculously,
from windings of the D-particle world-lines in the eleventh dimension.

\medskip

The one-loop diagram can, in principle,  be obtained using  
covariant Feynman
rules by  summing over the  contributions of the component  fields  
circulating
in the loop --- the graviton, gravitino and third-rank  
antisymmetric tensor
fields.  Alternatively, it can be expressed in terms of on-shell  
superfields.
  In that case the dynamics is defined by superspace quantum  
mechanics with the
massless superparticle action which reduces in a fixed  
parameterization of the
world-line to \refs{\kallosh},
\eqn\superpar{S_{particle} ={1\over 2} \int d\tau G_{\mu\nu}  (\dot X^\mu  - i \bar
\Theta \Gamma^\mu \dot \Theta) (\dot X^\nu  - i \bar \Theta  
\Gamma^\nu \dot
\Theta),}
where $\Theta$ is a 32 component $SO(10,1)$ spinor, $\mu =1,  
\cdots,11$    and the
reparameterization constraint requires the action density to vanish  
on physical
states.
  For present  purposes it will be sufficient to limit consideration to  
processes in
which the external states do not carry  momentum in the eleventh  
dimension and
which are also not polarized in that direction although these are
not essential conditions.   This loop
amplitude can be calculated by making use of the light-cone  
description of the
super-particle in which the vertex operator for a graviton has the form,
\eqn\vertgrav{V^{(r)}({\zeta^{(r)}, \tau^{(r)}}) = \zeta^{(r) ik}  
(\dot X^i - {1\over 4 p^+}
\theta \gamma^{ij} \theta k^{(r)}_j)(\dot X^k - {1\over 4 p^+} \theta  
\gamma^{kl} \theta
k^{(r)} _l) e^{ik^{(r)}\cdot X},}
where $i,j,\cdots = 3, \cdots, 11$, $\zeta^{(r)}_{\mu\nu}$ is the  
graviton wave
function with momentum  $k_\mu^{(r)}$   (where $(k^{(r)})^2 =0=
k^{(r)\mu}\zeta^{(r)}_{\mu\nu}$),  $\theta^a$ is a $SO(9)$ spinor    
in the
light-cone gauge defined by $\Gamma^+ \Theta = 0$ and  $X^+ = p^+  
\tau $ (where
$V^{\pm}  \equiv V^1 \pm V^2$ with timelike $V^1$).   This vertex  
operator is
attached at a point $\tau^{(r)}$ on the world-line and  is  defined  
in a frame
in which $k^{(r) +} =0$, $\zeta^{(r) +\mu } =0$.   In a canonical  
treatment of
this system the equations of motion determine that $X^i = p^i \tau  
+ x^i$ and
$\theta^a = S^a / \sqrt{p^+} $ and the (anti)commutation relations are $[p^i,  
x^j] = -i
\delta^{ij}$, $\{S^a,  S^b\} =  \delta^{ab}$ (just  
as with the
zero mode components of the corresponding relations in the  
ten-dimensional type
IIA superstring theory).

The loop amplitude reduced to $(11-n)$ dimensions by  
compactification on an
$n$-dimensional torus, $T^n$,  has the form,
\eqn\oneloop{\eqalign{A^{(n)}_4 & =  {1\over \pi^{5/2}   
\calV_n}
{\rm Tr} \int  d^{11-n} {\bf p}  \int_0^\infty  {d\tau \over \tau}
\left(\prod_{r=1}^4 \int_0^\tau d\tau^{(r)} V^{(r)}\right)  
\sum_{\{l_I\} }
e^{-\tau \left({\bf p}^2 +    G^{{(n)} IJ} l_I l_J \right)}\cr
  & =  {1\over \pi^{{n\over 2} -3} \calV_n} \tilde K   
\int_0^\infty d\tau
\tau^{n/2 - 13/2}  \sum_{\{l_I\} } e^{- \tau G^{(n)IJ} l_I  
l_J}\int_0^\tau
\prod_{r=1}^4 d\tau^{(r)} F(\{k^{(r)}, \tau^{(r)}\}) , \cr}}
where ${\bf p} = p^i$ is the $(11-n)$-dimensional loop momentum  
transverse to
the  compact directions and $G^{(n)}_{IJ}$ ($I,J = 1, \cdots n$) is  
the metric
on $T^n$ which has volume $\calV_n= \sqrt{\det G^{(n)}}$.    The  
kinematic
factor, $\tilde K$, in the second line involves eight powers of the  
external
momenta and follows from the trace over the components of $S^a$.  It may also be written as
\eqn\intferm{\tilde K \sim \int d^{16} \eta \prod_{r=1}^4\left(
\zeta^{(r)}_{\mu_r\omega_r}k^{(r)}_{\nu_r} k^{(r)}_{\tau_r}\bar\eta
\Gamma^{\mu_r\nu_r \rho_r}\eta  \bar\eta \Gamma^{\omega_r  
\tau_r}_{\ \
\ \ \rho_r} \eta\right) ,}
where $\eta$ is a chiral $SO(9,1)$ Grassmann spinor.
The overall normalization will be chosen so that $\tilde K$ is the  
linearized
approximation to
 \eqn\tensterm{\Rfour\equiv t^{\mu_1\dots \mu_8} t_{\nu_1\dots \nu_8}
R^{\nu_1\nu_2}_{\mu_1\mu_2} \cdots R^{\nu_7\nu_8}_{\mu_7\mu_8}}
where the tensor $t_8^{\mu_1\cdots \mu_8}$ was defined in
\refs{\greenschwarz}.
The function $F$ is  a simple function of the external momenta.   
Since we are
interested here in the leading term in the low-energy limit (the   
$\Rfour$
term)  we can set the momenta $k^{(r)}$   to zero in the integrand   
so that
$\int \prod d\tau^{(r)} F$ is replaced by $\tau^4$ giving
\eqn\oneloop{A^{(n)}_4   =  {\pi^{3/2}\over  \calV_n}  
\tilde K
\int_0^\infty d\tau \tau^{n/2 - 5/2}  \sum_{\{l_I\} } e^{-\pi \tau  
G^{(n)IJ} l_I
l_J}.}
Though this expression was obtained in a special frame we know   
that there is
an $(11-n)$-dimensional covariant extension  (including the case  
$n=0$) that
would follow directly from the covariant Feynman rules and should  
be easy to
check by an explicitly calculation using the component form of the  
supergravity
field theory action.

The expression for $A^{(n)}_4$ will  contribute  to the  
$\Rfour$ terms
in the effective action for M-theory compactified on $T^n$.
In order to  determine the dependence of the amplitude on the  
geometry of the
torus on which it is compactified it will be important to express   
$A_4^{(n)}$ in
terms of the winding of the loop around $T^n$.  This could be  
obtained directly
from the definition of the loop amplitude as a functional integral or by
performing a Poisson summation on the $n$ integers, $l_i$, which  
amounts to
inverting the metric in \oneloop.  The result is
\eqn\oneagain{A^{(n)}_4    = \pi^{3/2} \tilde K  \int_0^\infty d\hat\tau  
\hat\tau^{
1/2}  \sum_{\{\hl_I\}} e^{-\pi \hat \tau G_{IJ} \hl_I \hl_J} ,}
where $\hat \tau = \tau^{-1} $.

The ultraviolet divergence of eleven-dimensional supergravity comes  
from the
zero winding number term, $\{\hl_I\} = 0$,  in the limit that the loop 
shrinks to a  point ($\hat
\tau \to \infty$).    We will formally  write this divergent term as the
ill-defined expression, $C\equiv \int d\hat \tau \hat\tau^{  1/2}  
$.\foot{The presence of a cubically divergent $\Rfour$ term 
in eleven-dimensional supergravity was first suggested in \tseytlin.} 
 The fact that the one-loop supergravity amplitude is infinite 
 is  a signal that   point-particle dynamics alone cannot  
determine the
short-distance physics of M-theory.   A microscopic theory --- 
such as  Matrix theory \refs{\banksetal} --- should 
 determine the correct finite value of $C$.
This is somewhat  analogous to the way in which divergent loop amplitudes in
 ten-dimensional super Yang--Mills  are regularized by  
ten-dimensional string theory (for example, the $F^4$ terms 
 in the effective action of the heterotic and open 
string theories \refs{\backir,\greeks}).
Indeed,   we
will soon see that consistency with the duality symmetries of  
string theory
together with the assumption that the eleven-dimensional theory can  
be obtained
as a limit of the lower-dimensional theories determines the precise finite
renormalized value for the constant, $C$, 
that is also consistent with eleven-dimensional supersymmetry.     
It is a challenge to Matrix
theory to reproduce this number.

In the following we will associate   the  integer   $\hl_r$ with   
the winding
number of the loop around  a compact dimension of   circumference  
$R_{12-r}$
(for $r \ge 1$).
If a single direction is compactified  on a circle of circumference
$R_{11}\equiv \calV_1$, the loop can be expressed as a sum over the  
winding
number of the euclidean supergraviton world-line.   \eqn\oned{    
{1 \over \pi^{3/2}}A^{(1)}_4
=C   \tilde K+   2 \tilde K  
\int_0^\infty d\hat \tau \hat \tau^{ 1/2}   
\sum_{\hat l_1 >
0} e^{- \pi  \hat \tau\hat l_1^2 R_{11}^2} =C \tilde K +  \tilde K  \zeta(3)
{1\over \pi R_{11}^3}.}
Rather strikingly, the finite $R_{11}$-dependent term  gives a term  
in the
effective ten-dimensional action that is  precisely  that obtained in
\refs{\grisaru,\grosswitt} from the tree-level  IIA string theory  
(here written in the  M-theory frame).      
Although  the
regularized constant, $C$, is still undetermined, we will see later  
that it
must be set  equal to the coefficient of the one-loop $\Rfour$ term  
of the low
energy effective action of  string theory.  The absence of any further
perturbative or nonperturbative terms is in accord with the  
conjectures in
\refs{\greengut,\greenhove}.

Compactification on a torus ($n=2$) gives a richer structure.  In  
this case the
 one-loop amplitude has the form
 \eqn\torusfin{\eqalign{{1\over \pi^{3/2}}\calV_2 A^{(2)}_4 &
 = \calV_2 C\tilde K
+{\calV_2^{-1/2}}\tilde K  \sum_{(\hl_1,\hl_2)\ne (0,0)}   \int d\hat\tau  \hat
\tau^{1/2}e^{-\pi \hat\tau{1\over  \Omega_2}   |\hl_1 + \hl_2 \Omega|^2  
} \cr &
= \calV_2 C \tilde K + {1 \over 2\pi  } \calV_2^{-1/2}\tilde K
\sum_{(\hl_1,\hl_2)\ne (0,0)}  {\Omega_2^{3/2} \over | \hl_1 + \hl_2
\Omega|^3}
\cr & = {1\over 2\pi} \tilde K \left(2\pi C \calV_2 + \calV_2^{-1/2} 
f(\Omega,\bar \Omega)
\right) ,\cr}}
where   the divergent  zero winding term, $\hl_1=\hl_2=0 $   has  
again been separated
from the terms with non-zero winding.
The function $f$  in this expression is precisely the  
(finite) $\Omega$-dependent term in \hdef.
 In particular, in the limit $\calV_2\to 0$ M-theory should reduce  
to type IIB
superstring theory in ten dimensions \refs{\aspinwall,\schwarza} with the
complex scalar field,  $\rho \equiv C^{(0)} + i e^{-\phi^B}$,   
identified with
$\Omega$ (where $C^{(0)}$ is the \RR\ scalar and $\phi^B$ is the  
IIB dilaton).
  More precisely, the correspondence between the parameters in   
M-theory  and
in  IIB is,
\eqn\rad{\calV_2 \equiv  R_{10} R_{11} =  e^ {{1\over 3}\phi^B}r_{B  
}^{-{4\over
3}}, \qquad \Omega_2  \equiv  {R_{10} \over R_{11}} = e^{-\phi^B} }
(where $r_{B }$ is the radius of the tenth  dimension expressed in  
the IIB
sigma-model frame).
Using the fact that $\sqrt{G^{(9)}} (\calV_2)^{-{1\over 2}} \Rfour  =
\sqrt{g^{B(9)}}  r_B \Rfour$ (where $g^{B(d)}$ denotes 
the determinant of the IIB sigma-model
metric in $d$ dimensions)  we see that \torusfin\  leads, in  the
ten-dimensional IIB limit ($r_B \to \infty$),  to the expression  
suggested in
\refs{\greengut}.   This has the property that, when expanded in  
perturbation
theory ($e^{-\phi^B}  = \Omega_2 \to \infty$),   it  exactly  
reproduces both
the tree-level and one-loop $\Rfour$ terms of the type IIB theory  
as well as an
infinite series of D-instanton terms
\refs{\greengut,\greenhove}.    Importantly, the divergent term in 
\torusfin\ is proportional to $\calV_2$ and does not contribute in the 
limit of relevance to ten-dimensional type IIB -- thus the eleven-dimensional 
one-loop calculation reproduces the complete, finite, 
$\Rfour$ effective action in the type IIB theory.

As before, the  coefficient of the tree-level term  in the  type IIB
superstring perturbation theory is reproduced by the configurations with
$\hl_2  = 0$,   in which the particle in the loop winds around the  
eleventh
dimension but not the tenth (obviously there is a symmetry under the
interchange of these directions so we could equally well consider   
the terms
with $\hl_1 =0$).   In order to expand  \torusfin\ systematically  
for large  $\Omega_2$ it is necessary to undo the Poisson summation  
on $\hl_1$
for the terms with $\hl_2 \ne 0$.
These terms are then expressed as a sum of multiply-wound D-particle
world-lines where the winding number is $\hl_2$ and the D-particle  
charge is
the Kaluza--Klein charge, $l_1$.  In the limit  $\calV_2\to 0$ the  
 terms with
$l_1 =0$  reproduce the one-loop  $\Rfour$ term of    
ten-dimensional type IIB
while  the  $l_1\ne 0$ terms give the  contribution  of the sum of
D-instantons.  The precise contribution due to  the world-line of a  
particular
wrapped massive  D-particle (of mass $l_1$ and winding $\hat l_2$) to this
instanton sum is  identical to that obtained by considering     
semiclassical
quantization of four-graviton scattering in this background.   
Supersymmetry
causes all quantum corrections to vanish.
The additional fact that the one-loop string theory result is  
equivalent to the
sum of windings of a massless D-particle (the supergraviton) is notable
\refs{\greenhove}.  From the point of view of  
the string calculation this term arises from 
wrapping the string world-sheet in a
degenerate manner around a circle.   

We can now use  the additional constraint of T-duality to pin down  
the precise
value of $C$. This is determined by recalling that the one-loop  
terms in both
the IIA and IIB theories are invariant under inversion of the  
circumference,
$r_A \leftrightarrow  r_B^{-1}$.   This equates the coefficients of the
$\calV_2$ term and the $\Omega_2^{1/2} \calV_2^{-1/2}$ terms in  
\torusfin, and the result is that the  
coefficient $C$
must be set equal to the particular value,
\eqn\cdef{C= {\pi\over 3}. }

The fact that the modular function in    \hdef\ is a Maass wave  
form satisfying
\fdef\ is easily deduced from the integral representation, \torusfin.
Developing a geometrical understanding of the  origin  of this  
equation would
be of interest.

Upon compactifying on $T^3$ new issues arise.   The full U-duality group  
is $\Sl3 \times \Sl2$.  The  seven moduli  consist  of the
six moduli 
associated with the
three-torus and  $C^{(3)}_{123}$, the component of the antisymmetric
three-form on the torus.  The latter  couples to the euclidean  
three-volume of the
M-theory two-brane which can wrap around $T^3$.   The perturbative
eleven-dimensional one-loop  expression can be expected to reproduce the
effects of the Kaluza--Klein modes  but not  of the wrapped Membrane
world-volume.  However, these wrapped Membrane effects will be determined in
 the following by imposing U-duality and making use of  the one-loop
 results for 
type II string theory compactified on $T^2$ \refs{\greenhove}.
We will write the complete four-graviton amplitude as
\eqn\compamp{\calV_3 A_4^{(3)} =\pi^{3/2}  \tilde K H,}
where the scalar function $H$ depends on the seven moduli fields. There 
are several distinct classes of terms that will make
separate contributions to the complete function $H=\sum_i H_i$.

The effects of the Kaluza--Klein modes are obtained from \oneagain\
with $n=3$.  In order to compare with string theory on $T^2$ we will
choose $R_{11}$ to be the special M-theory direction so that $R_{11} =
e^{2\phi^A/3}$, where $\phi^A$ is the IIA dilaton (although the
expression obviously has complete symmetry between all three compact
directions). The sums over windings will be divided into various groups
of  terms.    Firstly,
there is the term with zero winding in all directions 
which is again divergent but  will be set equal to the regularised value  
given by  $C$ in \cdef, which implies
\eqn\divterms{ H_1 =  {\pi \over 3} \calV_3 \equiv {\pi \over 3}  T_2,}
where $T_2$ is the imaginary part of the K{\"a}hler structure of $T^2$.
The sum over $\hat l_1 \ne 0$ with $\hat l_2 = \hat l_3
=0$ once again leads to the correct tree-level string contribution
proportional to $\zeta(3)$,
\eqn\treeag{H_2 = \zeta(3) {1\over \pi R_{11}^3}= \zeta(3){1\over \pi}
e^{- 2\phi^A}.}

 The remaining sum is over all values of 
$\hat l_1$,  $\hat l_2$ and $\hat l_3$ excluding the $\hl_2 = \hl_3=0$ terms.
This is  usefully reexpressed by converting the $\hat
l_1$ sum to a sum over Kaluza--Klein modes by a Poisson resummation. The sum
of these terms  is
\eqn\tornon{
\eqalign{
H_3+H_4 =&  
 {\sqrt{\det G \over G_{11}}} \sum_{(\hl_2,\hl_3)\ne (0,0)}
\sum_{l_1} \int_0^\infty d\hat\tau \exp\left[ 2\pi i l_1 \hl_2 {G_{12}\over
G_{11}} + 2\pi i l_1 \hl_3 {G_{13} \over G_{11}} \right.  \cr & \left. 
- \pi l_1^2 {1\over \hat \tau G_{11}} - \pi\hat\tau
\left(\hl_2^2(G_{22} - {G_{12}^2\over G_{11}}) + \hl_3^2 (G_{33} -
{G_{13}^2 \over G_{11}}) +2 (G_{23} - {G_{12}G_{13} \over G_{11}})\hl_2
\hl_3 \right) \right] \cr
 &  = T_2
\sum_{(\hl_2,\hl_3)\ne (0,0)} \sum_{l_1} \int_0^\infty d\hat \tau \exp\left[ -
\pi l_1^2 e^{-2\phi^A} {1\over \hat \tau} + 2\pi i l_1 \hl_i A^{(i)} -
\pi \hat \tau \hl_i \hl_j g^A_{ij} \right] .\cr}}
 In this expression
$G_{ij}$ is the metric on $T^3$ in M-theory coordinates with the
convention that $i = 12 - \mu$ ($i = 1,2,3$) and the components of the
IIA string sigma-model metric on the two-torus are given by
\eqn\gadef{g^A_{ij} = R_{11} \left(G_{ij} - {G_{1i} G_{1j} \over
G_{11}} \right),} where $i,j =2,3$.  The components of the \RR\
one-form potentials in the directions of the two-torus in \tornon\ are
defined by 
\eqn\oneform{A^{(i)} = {G_{1i} \over G_{11}}.}  
The expression \tornon\ depends on the \RR\ one-form, 
the complex structure of the two-torus, \eqn\compstr{U = {1 \over
g^A_{22}}(g^A_{23} + i\sqrt {\det g^A})} 
and the combination $T_2 e^{-2\phi^A}$.
But it does  {\it not} depend separately on  the K{\"a}hler structure,
\eqn\kahlers{T = B_{12} + i \sqrt {\det g^A} = C^{(3)}_{123} + i \calV_3,}
where $\calV_3 = R_9 R_{10} R_{11}$.    
In the last step we have used the usual identification of
the \NSNS\ two-form with the M-theory three-form, $C^{(3)}$, and the
fact that $r^A_2 r^A_3 = R_9 R_{10} R_{11}$, where $r^A_i$ is the
circumference of the dimension labelled $i$ in the IIA sigma-model
frame. 

The expression \tornon\ contains perturbative and non-perturbative
contributions to the $\Rfour$ term in the IIA effective action.  The
perturbative term is obtained by setting $l_1=0$.  The resulting
double sum over $\hl_2$ and $\hl_3$ is logarithmically divergent, just as
in the analogous problem considered in  \refs{\dkl}.   This is a reflection
of the fact that the one-loop diagram in eight-dimensional supergravity is 
logarithmically divergent.   As in \refs{\dkl,\greenhove}, 
this divergence may be
regularized in a unique manner that is consistent with modular invariance 
by adding a term,  
$\Sigma = \ln (T_2 U_2/\Lambda^2)$, giving, 
\eqn\louisa{
\eqalign{ H_3  =& 
 \sum_{(\hl_2, \hl_3)\not = (0,0) } {U_2\over |\hl_2+ \hl_3 
U|^2 } -
\ln(U_2 T_2/\Lambda^2)\cr
 =&-
\left[\ln(U_2|\eta(U)|^4) + \ln(T_2)\right],  }}
where $\Lambda^2$ is adjusted to cancel the divergence coming from the
sum.
  So we see that the piece of the perturbative
string theory one-loop amplitude that depends on $U$ 
is reproduced by configurations in which a massless
particle propagating in the loop has a world-line that winds around
the torus.   This is the generalization  
of the way in which the IIB one-loop term 
was reproduced earlier  by
windings of a massless D-particle around a circle.

The terms with $l_1 \ne 0$ in \tornon\ consist of a sum of
non-perturbative D-instanton contributions,
\eqn\nonpert{H_4 =
 2 U_2 \rho^A_2
\sum_{(\hl_2,\hl_3)\not=(0,0)\atop l_1\not = 0} {|l_1|\over |\hl_2+ \hl_3U|}
K_1\left(2\pi \rho_2^A | \hl_2+ \hl_3U| |l_1| \right) e^{2i\pi l_1
(\hl_2 A^{(1)} + \hl_3 A^{(2)})},  }
where $\rho_2^A = r_2^A e^{-\phi^A}$. 
Using the fact that $K_1(z)= \sqrt{\pi\over 2z} e^{-z} (1+o(1/z))$ for large $z$ we see that at weak  IIA coupling, $e^{-\phi^A } \to \infty$, these terms are exponentially suppressed.
The contribution of these instanton terms in the nine-dimensional case 
described earlier is obtained by letting $r_3^A \to \infty$.  In this case 
$U\to i\infty$ and only the $\hl_3 =0$ term in \nonpert\ survives. 
 The double sum over $l_1$ and $\hl_2$ becomes the nine-dimensional D-instanton sum contained in \torusfin\ which was explicitly given in \refs{\greenhove}.

So far we have ignored the contributions to the $\Rfour$ term  arising 
from configurations in which the world-volume of the M-theory Membrane is 
wrapped around $T^3$.  
 Such contributions are obviously not contained in the one-loop $D=11$  
supergravity
 amplitude.  As with the contributions that came from circulating D-particles,
the configurations that contribute to the $\Rfour$ term  are described by the 
multiple windings of world-lines of nine-dimensional BPS states in ultra-short 
(256-dimensional)
multiplets.  Recall that these nine-dimensional states
are winding states of fundamental 
strings with no momentum or oscillator excitations which 
 are configurations of the wrapped M-theory Membrane with
 no Kaluza--Klein excitations.  Such contributions are therefore labelled by
two integers and depend only on the volume
 of the three-torus, $\det G$, and on $C^{(3)}$ but are independent of the 
other 
five components of the metric (i.e., they depend only on $T$ and $\bar T$). 
These  configurations of the IIA string world-sheet are just  those 
that enter the functional integral for the $\Rfour$ 
term at  one loop in string perturbation theory.
Indeed, as explained in \refs{\greenhove} 
(and in an analogous problem in \refs{\backir,\greeks}),
 the piece of the one-loop string amplitude that depends on $T$ and $\bar T$ 
is given by a sum over non-degenerate wrapped world-sheets and contributes,
\eqn\louisb{H_5  = 2 \sum_{m, n>0} {1\over n} 
\left(e^{2\pi i mn T} + e^{-2\pi i  mn \bar T}\right)
= -  \left[
\ln(|\eta(T)|^4) +{\pi\over 3} T_2\right],}
where $m,n$ are the integers that label the windings of the world-sheet. 
The sum of $H_1$, $H_3$ and $H_5$ reproduces the full one-loop perturbative 
string theory result.  Applying  
 T-duality in one of the toroidal directions transforms this into the one-loop
 term of the IIB theory.   The complete  non-perturbative structure of the
 ten-dimensional $\Rfour$ terms of the IIB theory can then be recovered
 using the   series of dualities described in
 \refs{\greenhove}.

The total contribution to the $\Rfour$ terms in the eight-dimensional effective
 action is  given (in IIA string coordinates) by 
\eqn\eighttot{S_{R^4} \sim  \int d^8 x \sqrt{g^{A(8)}}\, r^A_2 r^A_3\, H \, \Rfour,}
where $H =  \sum_{i=1}^5 H_i$ and we have ignored an overall constant.
  This expression is invariant under the requisite $\Sl3 \otimes \Sl2$
U- duality symmetry.  The particle winding numbers $(\hl_1,\hl_2,\hl_3)$
 transform as a ${\bf 3}$ of $\Sl3$ while the windings of the Membrane $(m,n)$
transform as a ${\bf 2}$ of $\Sl2$.  The decoupling of the two factors in the
U-duality group arises from the fact that the ultra-short BPS states in nine
dimensions do not contain both a  wrapped Membrane and Kaluza--Klein charges.
Compactification on $T^4$ to seven dimensions is more complicated  since the 
U-duality group is $\Sl5$, which is not a product of two factors.  The
 $\Rfour$ terms in this case depends on the  BPS spectrum in eight dimensions,
which was  discussed in \refs{\distler}.

\bigskip
In this paper we have studied properties of the one-loop  
amplitude in
eleven-dimensional supergravity compactified on tori to lower  
dimensions.  Upon 
compactifying to 
nine dimensions on $T^2$ this amplitude reproduces the 
complete perturbative and non-perturbative $\Rfour$ terms in the effective
actions for the corresponding string theories if the ulta-violet divergence
is chosen to have a particular finite regularized value (a value that 
can presumably be derived from Matrix Theory).  
This  value is also  in agreement with that obtained by supersymmetry 
 which  
relates it
to  the $C^{(3)} \wedge X_8$ term \refs{\greenhove}.  
In the  
limit in which the two-torus has zero volume, $\calV_2 \to 0$,  the
regularized term  does not contribute and the  complete 
$\Rfour$ term of the ten-dimensional IIB theory is reproduced precisely
by the one-loop supergravity calculation.  
 It is noteworthy that the $\Rfour$ terms in the IIB theory only 
 get string-theory perturbative
contributions at  tree-level and one loop, in addition to the 
non-perturbative  D-instanton contributions.   
This is tantalizingly similar to the structure
of the $F^2$ terms in $N=2$ super Yang--Mills theory in $D=4$ dimensions.

Upon  compactification to eight dimensions
on  $T^3$ the one-loop eleven-dimensional 
supergravity amplitude reproduces the $\Sl3$-symmetric piece of 
the $\Rfour$ term that is associated with Kaluza--Klein instantons.  The
remaining piece that arises from the wrapped Membrane is uniquely determined
by consistency with the T-duality that relates the IIA and IIB theories,
together with one-loop string perturbation theory.
 We have not addressed  the new issues that arise in compactification on
 manifolds of non-trivial holonomy or compactification to lower
 dimensions. For example, compactification on $T^6$ requires considerations of
 the wrapped world-volume of the M-theory five-brane.

In addition to the $R^4$ terms considered here there are many other terms of
the same dimension involving the other fields of
 ten-dimensional string theory and  M-theory.  In the language of type
IIB supergravity some of these terms conserve the  R-symmetry charge  
(as with  the $R^4$ 
term) while some of them  violate it in a manner consistent with the instanton
effects (such as the $\lambda^{16}$ term described in \refs{\greengut}).

Since there is no scalar field there is no possibility of a well-defined
perturbation expansion in powers of a   small coupling constant in
eleven-dimensional supergravity. Fortuitously, 
the relation of the $\Rfour$ term 
to the eleven-form, $C^{(3)} \wedge X_8$, via supersymmetry, implies
that the one-loop expression  is exact with no  
corrections
from higher-loop diagrams (since the normalization of the eleven-form is 
fixed by anomaly cancellation).  This adds to the ever-increasing body  
of evidence
that the constraints of maximal supergravity are profoundly  restrictive.

\vskip 0.5cm
{\it Acknowledgements:} We wish to acknowledge   EC support under the
Human Capital and Mobility programme.  
MBG is grateful to the University of Paris VI where this work was completed.

\listrefs
\end